\documentclass[aps,reprint,twocolumn,floatfix]{revtex4-2}
\usepackage{amssymb}
\usepackage{graphicx}
\usepackage{amsmath}
\usepackage{epsfig}
\usepackage{array}
\usepackage{multirow}
\usepackage{color}
\usepackage{esint}
\usepackage{bm}
\usepackage{bbm}
\usepackage{epstopdf}
\usepackage{mathtools}
\usepackage{soul}
\usepackage[dvipsnames]{xcolor}
\usepackage{tikz}
\usepackage{tabularx}

\usepackage[
    bookmarks=true,
    pdftex,
    colorlinks=true,
    urlcolor=blue,
    linkcolor=black,
    citecolor=blue,
    breaklinks=true,
    hypertexnames=false
]{hyperref}
\usepackage{multibib}
\usepackage{physics}
\usepackage{makecell}
\usepackage{booktabs}

\begin{document}

\title{Observing Bosonic Enhancement and Suppression of Forward Light-Atom Scattering}

\author{Yufei Wang}
\affiliation{State Key Laboratory for Mesoscopic Physics and Frontiers Science Center for Nano-optoelectronics, School of Physics, Peking University, Beijing 100871, China}

\author{Yiheng Fan}
\affiliation{State Key Laboratory for Mesoscopic Physics and Frontiers Science Center for Nano-optoelectronics, School of Physics, Peking University, Beijing 100871, China}

\author{Bo Song}
\email{bsong@pku.edu.cn}
\affiliation{State Key Laboratory for Mesoscopic Physics and Frontiers Science Center for Nano-optoelectronics, School of Physics, Peking University, Beijing 100871, China}

\begin{abstract}
The interplay of light and atoms is essential for precisely probing quantum matter, revealing the collective phenomena and microscopic correlations. Here, we investigate the forward light scattering from ultracold bosons at small angles using interferometric detection. The scattering is precisely encoded in the interference between the reference and scattered fields in the Fourier plane, and the detection sensitivity is further improved by an adaptive intensity calibration. We observe enhanced scattering due to bosonic stimulation and its suppression by repulsive atomic interactions. Both the measured contrast and effective optical depth exhibit similar trends in the enhancement and suppression of light scattering. Our findings establish new opportunities for using forward scattering to probe correlations and characterize the properties of quantum many-body systems.
\end{abstract}

\maketitle

\section{Introduction}

Photon scattering is essential in resolving microscopic
structure, from probing crystal
structures to characterizing many-body systems of quantum gases~\cite{Bragg1913Structure,bloch2008many}.
In classical light scattering, photons scatter independently from individual particles, underlying a wide range of spectroscopic and imaging techniques,
including Rayleigh and Raman scattering~\cite{born1964principles,Devereaux2007Inelastic}.
However, as the temperature approaches zero,
scattering behaves collectively, in which quantum statistics and correlations
between atoms can play an important role.
For example, scattering can be enhanced by bosonic stimulation, whereas suppressed by Pauli exclusion or repulsive interactions.
These effects are analogous to photon bunching and antibunching in quantum optics, as revealed by the Hanbury-Brown-Twiss experiment~\cite{Hanbury1956Correlation,Glauber1963Quantum,Kimble1977Photon}.

Ultracold atoms provide a pristine platform for investigating collective light-atom scattering. Theoretical investigations of these phenomena date back more than three decades~\cite{Svistunov1990Resonance,Politzer1991Light,Javanainen1994Optical,
Morice1995Refractive,Javanainen1995OffResonance,Politzer1997Bose}, but they have only recently been observed through scattering from
ultracold Bose and Fermi gases~\cite{Deb2021Observation,Margalit2021Pauli,Sanner2021Pauli,Lu2023Bosonic, Konstantinou2026Suppression,Lu2026Suppression}. However, these experiments focus on scattering processes at large scattering angles, i.e., momentum transfer comparable to the photon momentum. In contrast, forward scattering at small momentum transfers remains largely unexplored.

Here, we realize a forward light-atom scattering and study how the enhancement and suppression of scattering from ultracold bosonic ytterbium (Yb) gas at small angles by detecting the interference between the incident probe and the scattered light fields in the Fourier plane.
Analogous to phase-contrast imaging~\cite{zernike1942phase,Andrews1996Direct,Meppelink2010Thermodynamics}, our interferometric scheme enhances the detection sensitivity to the scattering field, particularly for dense atomic clouds. The scattering strength is determined by both the atomic density profile and the correlations between atoms. In addition, we develop adaptive intensity calibration to improve the detection sensitivity. 

We investigate light scattering over a broad temperature range and characterize its dependence on the total phase-space density. Our forward-scattering measurements demonstrate an enhancement of light scattering due to bosonic stimulation, which is suppressed by atomic repulsive interactions, in agreement with theoretical simulations. Furthermore, measurements of the effective optical depth reveal a similar temperature dependence of the effective scattering cross section. Our findings reveal the important role of quantum statistics and many-body correlation in the light-atom scattering, paving the way for precision detection of quantum many-body systems. 

\section{Interferometric detection}

Our experiment begins with bosonic $^{174}\mathrm{Yb}$ atoms, of which ${}^{1}S_{0}$ electronic ground state has zero electronic angular momentum and the nucleus has zero nuclear spin. This provides an ideal platform for studying light scattering without the complications arising from Raman coupling between internal sublevels and spin exchange collisions~\cite{Lu2023Bosonic,Konstantinou2026Suppression}. A gas of $1.6\times10^{5}$ atoms at a temperature of $100\,\mathrm{nK}$ is prepared by evaporative cooling in a crossed optical dipole trap following an efficient two-color magneto-optical trap~\cite{li2025two,tao2026optical}.

\begin{figure}[tbp]
    \centering
    \includegraphics[width=\linewidth]{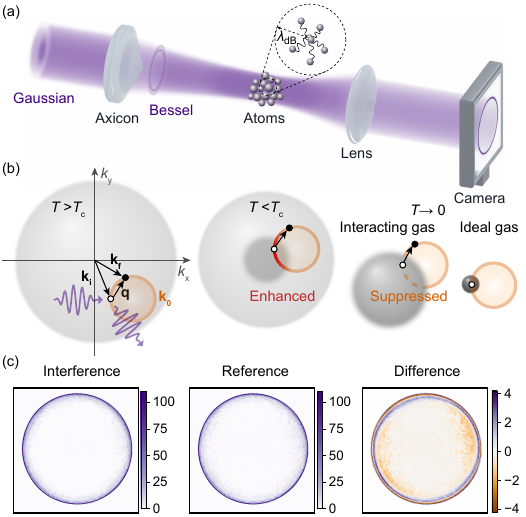}
    \caption{\textbf{Experimental setup for detecting bosonic enhancement and suppression.} (a) An annular probe beam, generated by passing a Gaussian beam through an axicon, is focused onto atoms. Cold atoms within a volume set by the thermal de Broglie wavelength $\lambda_{\rm dB}$ scatter collectively. The scattered light interferes with the probe field results in an interference pattern. (b) The enhancement and suppression of light-atom scattering at different temperatures. The probe light (purple) propagates along $k_{\rm x}$ and is scattered by the atoms, transferring momentum $\hbar q$, with the final atomic state lying on the orange ring of radius $k_{0}$, the wave number of the light. For thermal atoms, $T>T_\mathrm{c}$, atoms occupy large momentum states (light gray shadow regime) and bosonic enhancement is negligible (light orange line). For $T<T_c$, scattering between the condensate (dark gray shadow regime) and thermal cloud is enhanced (red line). Repulsive atomic interactions suppress bosonic enhancement by reducing density fluctuations. In the strongly repulsive interacting regime at $T=0$, interactions inhibit scattering into occupied states (dashed line), resembling Pauli blocking in Fermi gases. For an ideal Bose gas, the enhancement diverges as $q$ approaches 0 at finite temperatures due to the large occupation of low-momentum states, while it vanishes for nonzero $q$ at $T=0$. (c) Difference between the representative interference (with atoms) and reference images (without atoms) reveals clear interference fringes, demonstrating light scattering by atoms. To correct intensity fluctuations, we developed an adaptive intensity calibration method and see Supplemental Material for details.
    }
    \label{fig:Schematic}
\end{figure}

Figure~\ref{fig:Schematic} illustrates the light-atom scattering and detection setup. Light-atom scattering is probed by an annular beam generated by an axicon and focused onto cold atoms. At the position of the atoms, the focused beam has a beam waist of approximately $17~\mu\mathrm{m}$.  
The incident and scattered light are then collected by a lens with a focal length of $150~\mathrm{mm}$ and imaged onto a camera in the Fourier plane. The probe light field eventually forms a narrow annular shape with a beam waist of approximately $21~\mu\mathrm{m}$, corresponding to $\Delta k/k_0 \simeq 1.4\times10^{-4}$ in the Fourier plane, where $k_0=2\pi/\lambda$ is the wavenumber of the probe light and $\lambda$ is the wavelength. 
This narrow annular distribution provides a reference field for interference between the probe and scattered fields. Similar to a carrier wave in the radio frequency communication, the probe field acts as a reference field that interferes with the scattered light, enabling its detection with an enhanced signal-to-noise ratio.
The probe light has an intensity of
$I/I_{\mathrm{sat}}=9$ and is red-detuned by
$750\,\mathrm{MHz}$ from the
$^1S_0-{}^1P_1$ transition of Yb atoms at $398.9\,\mathrm{nm}$.

In the interferometric detection, we take two types of images, a reference image without atoms and an interference image with atoms. The interference image between the probe light field and the scattering field is similar to the probe reference image in Fig.~\ref{fig:Schematic}(c). However, the difference image between the interference and reference images, $I_{\mathrm{diff}}$, has a clear fringe pattern, reflecting the photon scattering from atoms. The small anisotropy arises from the nonspherical shape of the atomic cloud and the spatial profile of the probe light field at the atomic plane.

\section{light-atom scattering}

We conside the scattering process of off-resonant probe light from ultracold atoms. When a photon with an initial wavevector $\mathbf{k}_{\rm i,L}$ is scattered by an atom, its wavevector changes to the final wavevector $\mathbf{k}_{\rm f,L}$. Meanwhile, the scattering atom with initial momentum $\hbar\mathbf{k}_{\rm i}$ acquires a momentum transfer $\hbar\mathbf{q}$ and transits to a final momentum $\hbar\mathbf{k}_{\rm f}$, with $\mathbf{q}=\mathbf{k}_{\rm i}-\mathbf{k}_{\rm f}=\mathbf{k}_{\rm f,L}-\mathbf{k}_{\rm i,L}$. Since $|\mathbf{k}_{\rm i,L}|=|\mathbf{k}_{\rm f,L}|=k_0$, for any given $\mathbf{k}_{\rm i}$, the allowed $\mathbf{k}_{\rm f}$ values lie on the orange ring in momentum space in Fig.~\ref{fig:Schematic}(b). This allowed regime is also referred to as the Ewald sphere~\cite{born1964principles}. The scattering strength depends on the two-body correlation between $\mathbf{k}_{\rm i}$ and $\mathbf{k}_{\rm f}$, which is governed by quantum statistics, temperature and interatomic interactions.

The bosonic enhancement of the scattering process is proportional to the number of atoms in the final state. For thermal atoms $(T>T_\mathrm{c})$, the population is distributed over momentum states and bosonic enhancement is negligible. In contrast, for cold atoms $(T<T_\mathrm{c})$, occupation at ground state significantly increases the scattering between the condensate and the thermal cloud due to bosonic stimulation. The enhancement diverges as $q=|\mathbf q|\rightarrow0$. In the extreme limit $T\rightarrow0$, an ideal Bose gas is condensed into the zero-momentum state. Consequently, at finite $q$, it is progressively suppressed due to the vanishing occupation of the corresponding momentum states. In contrast, for a repulsively interacting gas at finite temperature, interactions suppress density fluctuations, the scattering is reduced but approaches a finite value in the $q\rightarrow0$, which gives rise to the compressibility of the gas. As $T\rightarrow0$, scattering is reduced as $q\rightarrow0$ but is nonzero as a consequence of the fluctuation-dissipation theorem~\cite{stringari2016bose}.

For an incident plane wave, each atom contributes a scattering amplitude with a position-dependent phase $e^{-i\mathbf q\cdot\mathbf r}$. The scattering intensity $I_{\rm sc}$ is proportional to the structure factor $S(\mathbf q)$, defined as follows~\cite{stringari2016bose},
\begin{equation}
    S(\mathbf q)=1+\frac{1}{N}\iint d^3\mathbf{r} d^3\mathbf{r'} [g^{(2)}(\mathbf{r},\mathbf{r'})-1]n(\mathbf{r})n(\mathbf{r'})e^{i\mathbf q \cdot (\mathbf r-\mathbf r')},
\end{equation}
where the first term 1 is contributed by
the scattering of single atoms, and $g^{(2)}(\mathbf{r},\mathbf{r'})$ is the second-order correlation function. We note that the effect of the density profile can also contribute to the scattering, which can be estimated using the density profile and see Supplementary Material for details.
The thermal de Broglie wavelength $\lambda_{\rm dB} =\sqrt{2\pi \hbar^2/(mk_{\rm B} T)}$ with the reduced Planck constant $\hbar$, mass of the atom $m$ and the Boltzmann constant $k_{\rm B}$. Meanwhile, $\ell_{\rm corr}$ is smaller than the cloud size $R$ in the experiment. The measured $q$ satisfies $q\ell_{\rm corr}\ll1$.
Therefore, for small $\mathbf{q}$ in the forward scattering, the structure factor reads, 
$S(\mathbf q)\approx 1 + N \tilde{g}^{(2)}$. Here $\tilde{g}^{(2)}=\iint d^3\mathbf{r} d^3\mathbf{r'} [g^{(2)}(\mathbf{r},\mathbf{r'})-1]n(\mathbf{r})n(\mathbf{r'})/N^2$ is the effective degree of second-order correlation. 
Therefore, the scattered field can be expressed as $E_{\rm sc}=\sqrt{I_{\rm sc}}$ in the forward scattering, yielding
\begin{equation}
    E_{\rm sc} \propto  \sqrt{1+ N\tilde{g}^{(2)}}.
\end{equation}
The total atom number $N$ characterizes the bosonic enhancement while $\tilde{g}^{(2)}$ accounts for modifications to this enhancement from the volume integration and atomic interactions. For ideal non-interacting Bose gas, the scattering is enhanced but the enhancement vanishes for $T\rightarrow0$; for strongly interacting bosons, scattering can be further suppressed.

\begin{figure}[tbp]
    \centering
    \includegraphics[width=\linewidth]{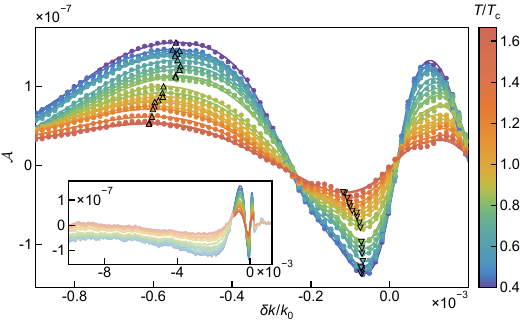}
    \caption{\textbf{Azimuthally averaged scattering signals at different diffraction angles for different temperatures.} The extracted scattering signal (circles) oscillates with angle due to phase modulation from the incident Bessel beam. The contrast of peak-to dip (triangles) reflects the scattering strength. The inset shows a wider range of diffraction angles and the solid lines are guides to the eye. Each curve is an average over 50 repetitions.}
    \label{fig:RadialA}
\end{figure}

The scattering can be extracted from the differential intensity $I_{\mathrm{diff}}=I_{\mathrm{interf}}-I_{\mathrm{ref}}
\approx 2 E_{\mathrm{ref}}E_{\mathrm{sc}}\cos\phi(k)$ with the incident probe field $E_{\mathrm{ref}}$ and intensity $I_{\mathrm{ref}}$. The azimuthally averaged difference image in Fig.~\ref{fig:RadialA} clearly shows the oscillatory interference fringes between the incident and scattering fields. $k=0$ is the center of the ring and $\delta k=k-k_{\rm r}$ represents the relative shift from the ring of radius $k_{\rm r} = 1.11\times 10^{-2} k_0$ in the Fourier plane in Fig.~\ref{fig:Schematic}. The phase profile of the Bessel beam introduces a spatially varying phase in the reference field, producing interference fringes around the reference-ring peak and determining the relative phase $\phi(k)$ between the incident and scattered fields.
Here we define the relative scattering strength $\mathcal{A}$, normalized by the probe intensity and atom number, as
\begin{equation}
    \mathcal{A}=\frac{\bar{I}_{\mathrm{diff}}}{2N\sqrt{\bar{I}_{\mathrm{ref}}}\,\max\left(\sqrt{\bar{I}_{\mathrm{ref}}}\right)}=\frac{E_{\rm sc}\cos\phi(k)}{N\max\left(\sqrt{\bar{I}_{\mathrm{ref}}}\right)}.
\end{equation}
The bar denotes the azimuthal average. The temperature dependence of the scattering reveals how quantum statistics and correlations determine the scattering process. The colder the temperature, the larger the contrast.
\begin{figure}[tbp]
    \centering \includegraphics[width=1\linewidth]{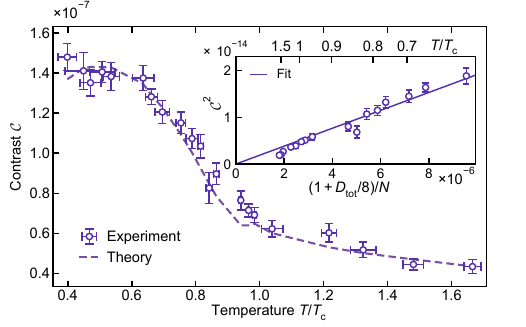}
    \caption{\textbf{Bosonic enhancement and its suppression at different temperatures.} Contrast $\mathcal{C}$ of $\mathcal{A}$ reflects the normalized scattering strength. For thermal atoms $T>T_{\rm c}$, $\mathcal{C}$ is small which is mainly attributed by the density effect and weak enhancement. As the temperature decreases below $T_{\rm c}$, $\mathcal{C}$ increases, indicating stronger enhancement. With further cooling, atomic interactions prevent such overlap and suppressing the increase of the enhancement. The measurement agrees with a theoretical simulation (dashed line) based on the experimental parameters and accounting for the effect of the atomic density profile. The inset shows that the squared contrast is approximately linearly proportional to the total phase-space density in both the thermal and near-transition regimes. The solid line represents a linear fit. Here each data point represents the average of 50 repetitions and the error bar denotes the standard deviation.}
\label{fig:contrast}
\end{figure}

Therefore, the scattering strength can be experimentally quantified by the contrast, 
\begin{equation}
    \mathcal{C}=\frac{\mathcal{A}_{\rm peak}-\mathcal{A}_{\rm dip}}{2}\propto \sqrt{\frac{1}{N}+\tilde{g}^{(2)}},
\end{equation}
where $\mathcal{A}_{\rm peak}$ and $\mathcal{A}_{\rm dip}$ denote the value of $\mathcal{A}$ at the peak (upward triangles) and dip (downward triangles), respectively in Fig.~\ref{fig:RadialA}. This contrast is proportional to the scattering field per atom. The contrast $\mathcal{C}$ increases with decreasing temperature in Fig.~\ref{fig:contrast}.
For thermal atoms $T > T_{\rm c}$, the scattering enhancement is weak, which is also observed in the large diffraction angle~\cite{Konstantinou2026Suppression,Lu2023Bosonic}. Around the BEC phase transition, the enhancement increases rapidly and becomes almost saturated when $T/T_{\rm c}$ approaches 0.6, where the condensate fraction is approximately 80\%. The overall trend of the contrast is captured by a theoretical simulation (dashed line) that uses the experimental parameters and accounts for the atomic density profile. We also find that the simulation is less sensitive to the wavefront at the atomic plane, and the uniform and Bessel beams yield similar results; see Supplemental Material for details.

The inset of Fig.~\ref{fig:contrast} shows the squared contrast, which quantifies the enhancement of photon scattering and probes the normalized structure factor $S(\mathbf{q})/N$. We find that, in both the thermal regime and the vicinity of the BEC transition, $\mathcal{C}^2$ exhibits a linear dependence on the total phase-space density $D_{\mathrm{tot}}=n \lambda^3_{\rm dB}$ with $n$ is the atomic peak density. In the experiment, $D_{\mathrm{tot}}$ is estimated using $D_{\mathrm{tot}}=\zeta(3)(T_c/T)^3$. This behavior can be understood from the low-momentum limit, i.e. the long-wavelength limit of the static structure factor, $\lim_{\mathbf{q}\rightarrow0}S(\mathbf{q})\approx 1+ D_{\rm tot}$/8 and see Supplemental Material for details. Furthermore, forward scattering provides a means of probing the isothermal compressibility of the atomic gas $\kappa_{\rm T}$,  $\lim_{\mathbf{q}\rightarrow0}S(\mathbf{q})=nk_{\rm B}T \kappa_{\rm T}$~\cite{stringari2016bose}. 
For the cold temperatures below $T_c$, the scattering enhances with decreasing temperature as the condensate fraction grows. However, with further cooling and increasing density, the enhancement is suppressed and eventually saturates, highlighting the role of repulsive interatomic interactions.

\begin{figure}[tbp]
    \centering
    \includegraphics[width=\linewidth]{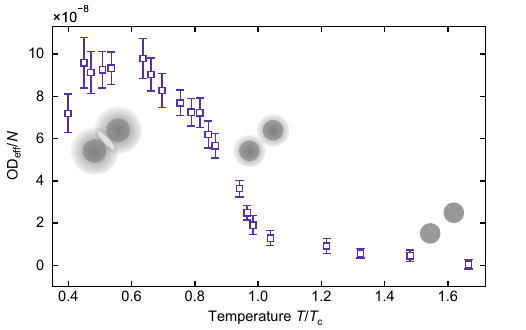}
    \caption{\textbf{Effective optical depth at different temperatures.} For thermal atoms, the scattering is weak for off-resonant light. And the enhancement is negligible (indicated by the gray ball). As the temperature decreases, the thermal de Broglie wavelength increases (indicated by ball boundaries), leading to stronger bosonic enhancement. With further cooling, atomic interactions prevent such overlap and suppress density fluctuations (indicated by the gaps between the balls), thereby suppressing the enhancement. Each data point represents the average of 50 repetitions and the errorbar denotes the standard deviation. 
    } 
    \label{fig:OD}
\end{figure}

The enhancement of scattering can also be interpreted in real space. Bosonic exchange enhances the probability of finding two particles at short separation, while repulsive interactions suppress this short-range bunching and attractive interactions can enhance it~\cite{Lu2023Bosonic,Konstantinou2026Suppression}. 
$\lambda_{\rm dB}^3$ defines a characteristic correlation volume. An atom has of order $n\lambda_{\rm dB}^3$ neighboring atoms within that volume, leading to collectively enhanced scattering. 
The intensity is integrated over a circular region, denoted by $\rm c$, with radius of $k_{\rm c} = 1.17\times 10^{-2} k_0$. We further measure the effective optical depth (OD), which characterizes the effective scattering cross section. It is defined as~\cite{Deb2021Observation}
\begin{equation}
    \mathrm{OD}_{\rm eff}=-\mathrm{ln}\left[ \frac{\int_{\rm c} I_{\mathrm{interf}}(\mathbf{k})d\mathbf{k}}{\int_{\rm c}I_{\mathrm{ref}}(\mathbf{k})d\mathbf{k}}\right].
\end{equation}
The effective OD per atom is plotted after subtracting the background measured for a hot atomic gas at $1\,\mu$K in which scattering is negligible. The resulting effective OD is shown in Fig.~\ref{fig:OD}. 

For thermal atoms at $T>T_{\rm c}$, scattering can be viewed as independent scattering from individual atoms (represented by the separated gray spheres), resulting in an effective OD close to zero and indicating negligible enhancement of the effective scattering cross section. As the temperature decreases below $T_{\rm c}$, the increasing thermal de Broglie wavelength leads to greater spatial overlap of the atomic wavefunctions, giving rise to enhanced bosonic scattering. Upon further cooling, however, interatomic interactions suppress density fluctuations (illustrated by the gaps between the spheres), thereby reducing the bosonic enhancement.
Both bosonic stimulation and interatomic interactions play roles in modifying the two-body correlation function and thus altering the scattering strength.

\section{Conclusion}

We have studied the forward scattering of light from ultracold bosons and its dependence on temperature. We detect the scattering via interferometric detection and extract the strength from interference between the scattering and reference fields in the Fourier plane, and further improve the sensitivity by an adaptive intensity calibration. We have observed the enhancement of scattering arising from the bosonic stimulation and suppression due to the repulsive interactions. Our findings highlight the role of spatial correlations in forward light-atom scattering and enable a route to precisely probe the density fluctuations and compressibility of quantum gases.

\section{Data Availability}

The data presented in this work are available on the Peking University
Open Research Data Platform.

\section{Acknowledgments}

This work was supported by the Quantum Science and Technology-National Science and Technology Major Project (2024ZD0301800), the Beijing Natural Science Foundation (Z240007), the National Natural Science Foundation of China (12374242), Scientific Research Innovation Capability Support Project for Young Faculty (ZY2025014) and SRAFS.

\bibliography{pci}

\section*{Supplemental Material}

\setcounter{equation}{0}
\setcounter{figure}{0}
\setcounter{table}{0}
\setcounter{section}{0}

\renewcommand{\thefigure}{S\arabic{figure}}
\renewcommand{\theequation}{S\arabic{equation}}
\renewcommand{\theHfigure}{S\thefigure}

\section{Adaptive intensity calibration}

In the experiment, the interference and reference images are acquired sequentially, making the subtraction susceptible to artifacts arising from shot-to-shot fluctuations in the probe intensity. Inspired by adaptive optics widely used in astronomy and microscopy~\cite{Hampson2021Adaptive}, we implement adaptive intensity calibration by splitting the incident beam with a beam splitter before the atomic sample and using the bypassing beam to monitor shot-to-shot probe intensity fluctuations.
\begin{figure}[htbp]
    \centering
    \includegraphics[width=\linewidth]{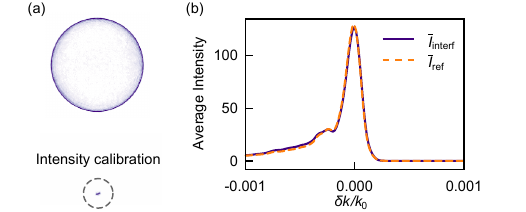}
    \caption{
        \textbf{Adaptive intensity calibration and azimuthal average.}
        (a) The reference light and the monitor spot (indicated by the dashed gray circle) are spatially separated. The integrated intensity of the monitor spot is used to correct intensity fluctuations. (b) Radially averaged
        intensity profiles of the light field after the adaptive intensity calibration.
    }
    \label{fig:adaptive}
\end{figure}
At the atomic plane, the monitoring spot is spatially separated from the atomic cloud and therefore does not scatter from the atoms. At the Fourier plane, the monitoring beam is focused onto a spot spatially separated from the Bessel ring, as shown in Fig.~\ref{fig:adaptive}(a).

We further quantity the performance of the adaptive intensity calibration by taking and analyzing 30 image pairs acquired without atoms. For each pair, we use the intensity correction factor $\eta=I_{\mathrm{interf}}^{\mathrm{spot}}/I_{\mathrm{ref}}^{\mathrm{spot}}$, where $I_{\mathrm{interf}}^{\mathrm{spot}}$ and $I_{\mathrm{ref}}^{\mathrm{spot}}$ are the monitor-spot intensities integrated in the interference and reference images, respectively. The raw and corrected radial profiles are calculated as
$\delta I_{\mathrm{raw}}=I_{\mathrm{interf}}-I_{\mathrm{ref}}$ and
$\delta I_{\mathrm{corr}}=I_{\mathrm{interf}}/\eta-I_{\mathrm{ref}}$,
respectively. We calculate the root-mean-square (RMS) residual among 30 pairs by choosing the signals within $3.4\times10^{-4} k_0$ on either side of the ring. The adaptive intensity calibration reduces the RMS residual from $1.43$ to $0.06$, corresponding to a $24$-fold suppression of the intensity fluctuation.

\section{Experimental details and data analysis}

\begin{figure}[htbp]
    \centering
    \includegraphics[width=\linewidth]
    {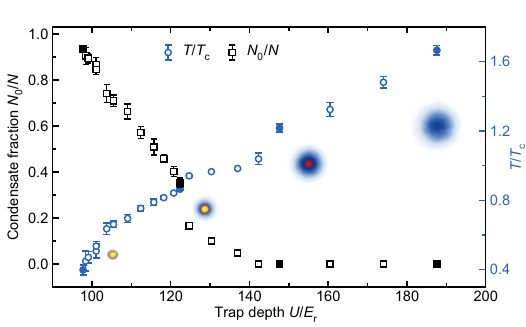}
    \caption{
    \textbf{Preparation of different temperatures by varying the dipole trap depth.}
    With decreasing the trap depth, the temperature (blue circles) decreases and the condensation fraction $N_0/N$ increases (black squares). Each data point represents an average of 20 repetitions and the error bar indicates the standard deviation. Insets show representative time-of-flight images corresponding to the filled symbols.
    }
    \label{fig:BEC}
\end{figure}

We prepare a cold gas of bosonic $^{174}$Yb atoms using two-color MOT cooling followed by forced evaporation in a crossed optical dipole trap~\cite{li2025two, tao2026optical}.
By controlling the evaporation duration and final trap depth $U$, a various of temperatures can be prepared as shown in Fig.~\ref{fig:BEC}. The resulting cold gas has $1.6\times10^{5}$ atoms at a temperature of $100\,\mathrm{nK}$, with a condensate fraction of up to $94\%$. In Fig.~\ref{fig:BEC}, for the condensed samples, $T/T_{\mathrm{c}}=(1-N_0/N)^{1/3}$ is calculated based on the fitted condensate fraction in the harmonic trap. For thermal samples with $T>T_{\mathrm{c}}$, the temperature $T$ is determined from a 2D Gaussian fit, while the critical temperature $T_{\mathrm{c}}$ is calculated using the trap frequencies inferred from the dipole-trap power. The condensate fraction $N_0/N$ is set to zero. Their error bars reflect only shot-to-shot temperature fluctuations, excluding $T_{\mathrm{c}}$ calibration uncertainty.

\begin{figure}[t]
    \centering
    \includegraphics[width=1\linewidth]
{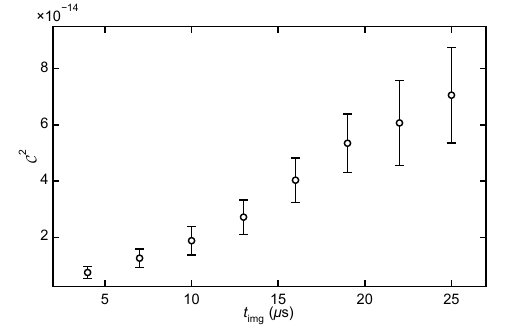}
    \caption{\textbf{Contrast squared as a function of the imaging exposure time.} The squared contrast, proportional to the scattered photon number, increases approximately linearly with exposure time in the fast imaging regime. Each data point is an average of 20 repetitions and the errorbar indicates the standard deviation.
    }
    \label{fig:timescan}
\end{figure}

We prepare an annular imaging beam by shining a Gaussian beam with a waist of approximately $1\,\mathrm{mm}$ onto an axicon. The probe light has a peak intensity of $I/I_{\mathrm{sat}}=9$ at the atomic cloud and is red-detuned by $\Delta=750\,\mathrm{MHz}$ from the $^1S_0 \rightarrow {}^1P_1$ transition of Yb atoms. The resulting beam is then focused onto the atoms using a lens of $300\,$mm. At the atomic plane, the central lobe has an average $1/e^2$ intensity radius of approximately $17\,\mu\mathrm{m}$. In the main text, the imaging pulse duration is set to $10\,\mu\mathrm{s}$. We further confirm that atomic motion is negligible and that the system remains within the linear-response regime, as shown in Fig.~\ref{fig:timescan}.
 
\section{Scattering simulation}

We consider atom-light scattering process in an interacting Bose gas. The photon with initial momentum $\hbar \mathbf k_{\mathrm{i,L}}$ is scattered by a single atom and obtains final momentum $\hbar \mathbf k_{\mathrm{f,L}}$. The cross section
of atom-light interaction with momentum transfer $\hbar \mathbf{q} = \hbar \mathbf k_{\mathrm{f,L}}-\hbar \mathbf k_{\mathrm{i,L}}$ is given by 
\begin{equation}
    \sigma(\mathbf{q}) = \sigma_0(\mathbf{q}) N S(\mathbf{q}),
\end{equation}
where $\sigma_0(\mathbf{q})$ is the scattering section of single atoms. The structure factor $S(\mathbf{q})$ is given by
the Fourier transform of the connected density-density correlation function,
\begin{align}
    S(\mathbf q)
    &=
    1+\frac{1}{N}\iint d^3\mathbf{r} d^3\mathbf{r'}   [\langle \psi^\dagger(\mathbf{r}) \psi^\dagger(\mathbf{r'}) \psi(\mathbf{r'})\psi(\mathbf{r}) \rangle \notag \\
    &-\langle \psi^\dagger(\mathbf{r}) \psi(\mathbf{r})\rangle \langle \psi^\dagger(\mathbf{r'}) \psi(\mathbf{r'})\rangle] e^{i\mathbf q \cdot (\mathbf r-\mathbf r')} \notag \\
    &=1+\frac{1}{N} \iint d^3\mathbf{r} d^3\mathbf{r'}  [g^{(2)}(\mathbf{r},\mathbf{r'})-1]n(\mathbf r) n(\mathbf r')e^{i\mathbf q \cdot (\mathbf r-\mathbf r')}.
\end{align}
Here $N$ is the total atom number. $\psi^\dagger(\mathbf{r})$ is the bosonic field operator and the two-body correlation function is given by $
\langle \psi^\dagger(\mathbf{r}) \psi^\dagger(\mathbf{r'}) \psi(\mathbf{r'})\psi(\mathbf{r}) \rangle = g^{(2)}(\mathbf{r},\mathbf{r'}) \langle \psi^\dagger(\mathbf{r}) \psi(\mathbf{r})\rangle \langle\psi^\dagger(\mathbf{r'})\psi(\mathbf{r'}) \rangle$. The atom density is $n(r)=\langle\psi^\dagger(\mathbf{r})\psi(\mathbf{r})\rangle$. $g^{(2)}(\mathbf{r},\mathbf{r'})$ is the degree of the second order correlation.

For thermal atoms in the dipole trap, the second-order correlation function $g^{(2)}(r)$ with $r=|\mathbf{r}-\mathbf{r'}|$ can be approximated as~\cite{Naraschewski1999Spatial}
\begin{equation}
	g^{(2)}(r)\approx 1+ \exp(-2\pi r^2/\lambda_{\rm dB}^2).
\end{equation}
Therefore, the structure factor of thermal atoms can be estimated as
\begin{align}
    S(q)\approx 1+D_{\rm tot}\exp(-q^2 \lambda^2_{\rm db}/8\pi)/8,
\end{align}
where the $D_{\rm tot}$ is the total phase-space density and $q=|\mathbf{q}|$. The denominator factor of $8$ arises from averaging over the harmonic trapping potential. In particular, $\lim_{q\rightarrow0} S(q)\approx 1+D_{\rm tot}/8$.

\begin{figure}[tbp]
	\centering
	\includegraphics[width=\linewidth]{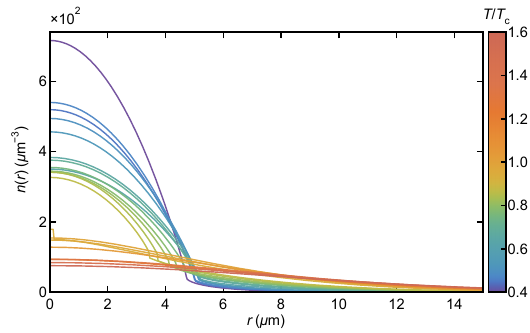}
	\caption{\textbf{Simulated density distribution in a harmonic trap.}
		The density distribution is calculated using the experimental parameters. Colors indicate different $T/T_{\rm c}$.}
	\label{fig:density}
\end{figure}

\begin{figure}[th]
	\centering
	\includegraphics[width=\linewidth]{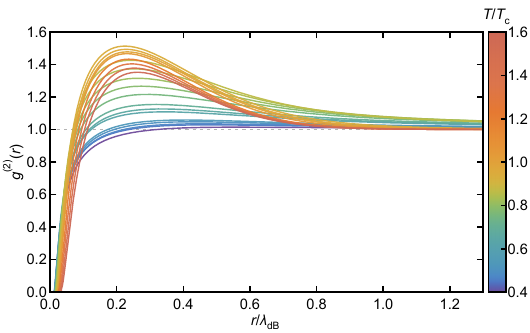}
	\caption{\textbf{Simulated correlation functions in a harmonic trap.}
		The trap-integrated correlation function is calculated using the experimental parameters. Colors indicate different $T/T_{\rm c}$.}
	\label{fig:correlation}
\end{figure}

We further estimate the second-order correlation function of interacting bosons in a harmonic trap. Firstly we calculate the density distribution using the self-consistent Hartree-Fock model~\cite{gerbier2004experimental}, as shown in Fig.~\ref{fig:density}. Then we follow the approach of Ref.~\cite{Naraschewski1999Spatial} to estimate the second order correlation. Figure.~\ref{fig:correlation} shows the averaged $g^{(2)}(r)$.

\begin{figure}[t]
	\centering
	\includegraphics[width=\linewidth]{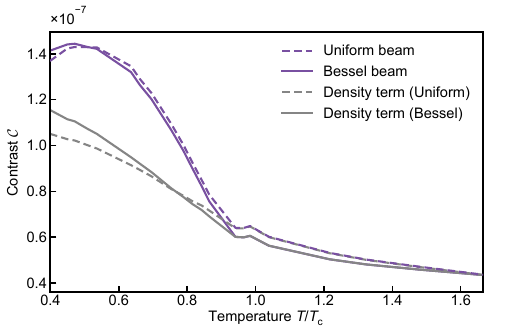}
	\caption{\textbf{Simulation of the scattering contrast.} The solid and dashed lines represent simulations for Bessel and plane-wave probe beams, respectively. The overall trend exhibits a weak dependence on the probe-beam profile. The gray lines show the contribution from the atomic density profile, calculated using the experimental parameters. The simulation indicates that the scattering enhancement is predominantly contributed by the regime below the critical temperature $T_{\rm c}$.}
	\label{fig:Ctheo}
\end{figure}

It is worth noting that the atomic density profile also affects the scattering amplitude and plays an important role in forward scattering. $S_{\rm d}(\mathbf{q})=|\int n(\mathbf{r})e^{i\mathbf{q} \cdot \mathbf{r}}d^3r|^2$~\cite{Lu2026Suppression}. This term contributes to coherent scattering of the atomic gas and changes the scattering intensity. In the simulation in Fig.~\ref{fig:contrast} in the main text, we account for this effect in the modification of the structure factor $S(q)$. We first extract the density-profile contribution factor by fitting with the measured contrast at $T/T_{\rm c}=1.66$, and then fit the residual using the structure factor, as shown in Fig.~\ref{fig:Ctheo}.

\end{document}